\documentclass[12pt]{iopart}

\begin{document}

\def\exd{{\rm d}}

\title[Challenges for String Cosmology]{Challenges for String Cosmology}

\author{C.~P.~Burgess${}^{1, 2}$ and Liam McAllister${}^3$}

\address{${}^1$Department of Physics \& Astronomy, McMaster University\\
1280 Main Street West, Hamilton ON, Canada.\\
\medskip
${}^2$Perimeter Institute for Theoretical Physics\\
31 Caroline Street North, Waterloo ON, Canada.\\
\medskip
${}^3$Department of Physics, Cornell University, Ithaca, NY 14853 USA.
}
%\ead{custserv@iop.org}
\begin{abstract}
We critically assess the twin prospects of describing the observed universe in string theory, and using cosmological experiments to probe string theory. For the purposes of this short review,
we focus on the limitations imposed by our incomplete understanding of string theory. After
presenting an array of significant obstacles, we indicate a few areas that may admit theoretical progress in the near future.

\end{abstract}

%\pacs{00.00, 20.00, 42.10}WeJOA, JOB?
\maketitle

\section{Introduction }

String theory has led to an impressive and growing array of insights on diverse problems in theoretical physics, but there has been comparatively little progress in understanding whether string theory actually describes our universe.  Unless the string scale is much closer to the TeV scale than to the Planck scale, decoupling makes it difficult to probe string theory in terrestrial experiments. Cosmological observations, however, provide a unique and powerful probe of ultraviolet physics, by revealing high-energy processes in the very early universe. Key aspects of cosmic history, from the origins of primordial perturbations, to the mechanism powering inflation, to the nature of the Big Bang singularity itself, can be sensitive to Planck-scale physics to varying degrees, and hence can illuminate -- albeit rather dimly -- the realm of quantum gravity.\footnote{Even
in cases that do not require a full ultraviolet completion, one
at least needs to specify the coefficients of Planck-suppressed operators in an effective theory below the Planck scale.}
This strongly motivates the use of observational cosmology as a constraint on string theory, and of string theory as the foundation for a more complete
description of the very early universe.

String theory has also had noteworthy success as a {\it{conceptual}} tool in cosmology, by providing toy models of cosmology in quantum gravity, and by inspiring new effective field theories admitting realistic cosmic histories. However, these motivations are thoroughly explored in many excellent reviews \cite{Reviews}, so we shall say little about them here. Instead, our goal is a critical assessment of the headier, and more distant, prospects of describing the observed universe in string theory, and using such a description to constrain string theory.

The reader can easily anticipate that given our very limited understanding of the vast space of solutions of string theory, testing the theory itself by any means is extremely difficult.  We do not disagree, but we note that much of the literature on string cosmology expresses far more optimism on this score.  Our goal is to sift out some of the most credible of these claims, and thus provide a necessarily personal view on what sorts of advances in string theory might eventually lead to meaningful cosmological constraints on the theory.  We hope that calling attention to some of the overlooked assumptions, significant obstacles, and traditional oversimplifications of the subject will help new entrants to the field to see where progress is most urgently needed.

As our intent is a critical, rather than encyclopedic, review, we do not attempt to cite all relevant original work, referring again to the reviews \cite{Reviews} for more extensive bibliographies.

\section{String cosmology in an ideal world}

Like the sirens of Greek mythology, string cosmology tends to draw intrepid explorers to founder on the reefs of calculational difficulty.
Two compelling goals motivate this quest.

The first of these is the tantalizing hope of illuminating physical processes at extremely high energies by observing the very early universe.
Our present understanding of cosmology is grounded in general relativity and quantum field theory, which have {\it{individually}} passed a remarkable panoply of experimental tests.
The primordial fluctuations seen in the cosmic microwave background radiation \cite{cmbinflation} provide an unprecedented experimental probe of the union of these two frameworks, and ultimately of the ultraviolet completion of gravity.
The broad outlines of a standard model of cosmic history are emerging: an early epoch of accelerated inflationary expansion \cite{inflation}, or perhaps of decelerating contraction \cite{contraction, Veneziano}, prepared a flat, homogeneous, isotropic universe  at the onset of the Hot Big Bang, and quantum fluctuations generated at this earlier time stretched to form the primordial perturbations.
Moreover, we are plausibly on the cusp of constraining the dynamics that generated the perturbations: impressive progress in observational cosmology is likely to rule out a large swath of phenomenological models in the next few years.

If only there were a precise, well-motivated theory that could decide what the dynamics of this very early universe was. String theory provides the (so far, unique)
%Liam June 10  framework  reduplication of this word
means of describing
%Liam June 10 in which to describe
this early epoch quantitatively, within a tightly constrained framework where sensible quantum calculations with gravity are possible.
In principle all
the marbles are on the table: one could hope to have a theory of initial conditions (how generic is the desired acceleration?) or of singularities (what happens when the universe is so small that quantum effects dominate?).
One
can hope to find a restrictive set of predictions from a small class of Lagrangians that allow a complete dynamical understanding of the intervening evolution, ideally leading to a collection of striking predictions that are clearly distinguishable from more prosaic models
built
directly in field theory. And since it is spacetime itself that is
in play, perhaps one can explain other important features of our late-time low-energy world, like why it appears to be four-dimensional.

The second great hope for string cosmology
%Liam June 10 also includes the cosmology
is tied to observations
of the much later, lower-energy universe. Although decoupling normally precludes inferring much about high-energy physics from very low-energy observations,
cosmological observations point towards some specific kinds of physics that appear not to decouple.
Most notably,
the present-day universe is undeniably very large, in a way that appears inconsistent with there being much energy density tied up in the vacuum, but in all our theories the vacuum energy density is notoriously sensitive to the details of
%Liam June 10  still confused about how to hyphenate the following line
very-high energy physics \cite{ccproblem}.
Unnatural situations like this, where low-energy observations disagree with generic expectations, can be the exceptions to the usual decoupling rule: understanding why the high-energy world produces such non-generic features at low energies can provide a rare insight into what the higher-energy physics really is.

The stakes are clearly very high, but we should not allow dreams of a complete and predictive theory of the early universe
to make us forget the intervening reefs, on which we might yet founder.

\section{String cosmology in the real world}

The problems start with the observation that theoretical predictions are useless for comparing with
%Liam June 10 observations
experiments unless they come with a reliable estimate of theoretical error. So a central part of making exciting, but testable, predictions is the comparatively dull exercise of carefully accounting for what controls the approximations made. This is particularly important for string cosmology,
whose signal virtue is the control string theory allows over the otherwise unruly high-energy interface between quantum effects and gravity.

Exact results have provided some of the crowning achievements of string theory, but it is difficult to apply these results to the time-dependent, non-supersymmetric backgrounds relevant for realistic cosmology. Essentially all of our knowledge of string theory in the cosmological setting is founded on approximations.  The elementary expansions are the weak coupling expansion, with string coupling $g_s \ll 1$, and the long-distance (or derivative) expansion, based on energies $E$ that are small relative to the string scale, $\alpha' E^2 \ll 1$. The expansions in these small quantities underpin the semiclassical reasoning using low-energy effective field theories upon which much of string-theoretic model building is based.

In practice, another kind of approximation
plays an important role in string cosmology.  A typical analysis collects `ingredients' that are understood to varying degrees in isolation, and assembles them in a single compactification with suitable cosmological properties (like a positive effective cosmological constant in four dimensions).  The approximation in question then amounts to the systematic incorporation of the interactions among these ingredients, {\em i.e.}\ expansion around the (fictitious) configuration in which the mutual interactions are neglected.
As an explicit example, consider
the KKLT \cite{KKLT} construction of de Sitter vacua in type IIB string theory, which
requires an elaborate assemblage of sources -- anti-D3-branes, nonperturbative effects on D7-branes, and fluxes -- that have not yet been combined in a single concrete solution admitting a systematic expansion. One therefore studies anti-D3-branes in
%Liam June 10 a
the noncompact Klebanov-Strassler
%Liam June 10
\cite{KS} solution, and approximates the corrections resulting from compactification as small.  One
%Liam June 10 determines  lots of word repetition fixed  here
computes the effective action for the compactification moduli in the approximation of weak warping and dilute fluxes, and argues that the strong warping found in examples of interest provides controllable corrections.  Finally, one %Liam June 10 determines the effective
obtains the gauge
theory on D7-branes in a (generally supersymmetric) flux compactification, and then appeals to strong gauge dynamics in the effective four-dimensional theory for the key ingredient effecting stabilization of the K\"ahler moduli. Determining the four-dimensional effective action
%Liam June 10 theory
therefore requires an intricate balancing of many approximations that are not strictly incompatible, but whose domains of overlapping validity can be slender and difficult to map. This fundamental requirement of remaining within the domain of validity of multiple expansions makes much of string cosmology rather baroque.

A further challenge is that in string theory the small control parameters arise as the values of fields, with the string coupling given by the dilaton, $g_s = e^{\phi}$, and $\alpha'$ compared in practice with derivatives of fields, like $\alpha' (\partial \phi)^2$ or
$\alpha' {\cal{R}}$, where ${\cal R}$
is the
Ricci scalar. In cosmology these fields evolve in time and space, and so care must be taken that they are never driven outside of the calculable regime. These are particularly dangerous issues in scenarios where the universe contracts, since the validity of the $\alpha'$ expansion is generally worse the smaller the universe is (about which we shall say more below).

On top of this comes the
%Liam June 10 complication of complexity:
challenge of complexity: string theory involves a great many fields, even at low energy. But gravity couples to all fields equally, and because cosmology's focus is the evolution of geometry, the energy tied up in the motion of {\em all} of these fields matters. This means that a convincing cosmology requires not only a calculation of how a particular set of fields evolve,
but also a demonstration that all of the other fields not included in
the calculation were not equally important. It is an oft-learned lesson in string cosmology that it is the fields {\em not} written down that can pose the biggest problems.  Historically this proved to be the most significant obstruction to progress: the plethora of moduli found in simple supersymmetric string compactifications
%Liam June 10 would
acquire a difficult-to-calculate scalar potential once supersymmetry
%Liam June 10 broke,
is broken, and this
%Liam June 10
potential is not negligible compared with the energies relevant for cosmology.
Thus, any
%Liam June 10 repetition  convincing
realistic string cosmology had to await the development of tools for understanding how to stabilize moduli.

\subsection{Stabilization of moduli}

The emergence in the past decade of methods for stabilizing moduli (for reviews
with original references, see \cite{modulusstabilizationreviews}) prepared the ground for a wide-ranging exploration of string cosmology.  Successful stabilization is the cornerstone of realistic cosmology in string theory, and limitations of present techniques for stabilizing moduli  are a key barrier to further developments of the subject.  In part because of the overlap with the contentious question of the landscape of string vacua, assessments of the subject vary considerably, from hypercritical to wildly optimistic.  The truth is less exciting: continued hard work by many authors has gradually enlarged our understanding of the space of stabilized vacua, but we are far from the level of understanding required to automate the search for vacua without moduli.

On the positive side, it is increasingly difficult to argue that string theory does not admit a vast number of stabilized vacua. Improved understanding of compactifications containing fluxes and D-branes
makes it clear that the required ingredients for some classes of stabilized compactifications are available.  Stabilization in AdS vacua, sometimes breaking supersymmetry, is very plausibly achievable by well-established mechanisms \cite{KKLT,LVS} in type IIB string theory.  AdS vacua
of type IIA string theory have also been obtained in various regimes ({\em cf.\ e.g.}\ \cite{DeWolfe:2005uu}).
In the heterotic string, there appear to be enough well-understood ingredients to give masses to the geometric moduli \cite{Anderson:2011cz}, though the requisite `racetrack' of large-rank gaugino condensates has
%Liam June 10 yet to be
%Liam June 10 demonstrated  (repetition)
not yet been constructed
in a compact model. In M-theory on
manifolds of $G_{2}$ holonomy, stabilization scenarios have been proposed ({\em cf.\ e.g.}\ \cite{Acharya:2006ia}), but control of the K\"ahler potential remains to be demonstrated.  Partial results are available at large volume \cite{Acharya:2005ez}, but singular loci are required to produce confining gauge theories, making it difficult to establish the detailed form of the K\"ahler potential.

Controllable breaking of supersymmetry in a stabilized Minkowski or de Sitter vacuum remains a more difficult problem. Although metastable breaking of global supersymmetry is comparatively well understood, and can be engineered in local models, coupling these field-theoretic mechanisms to gravity is not at all straightforward. This is because compactification promotes the control parameters of global supersymmetry to dynamical fields, creating new runaway directions that efficiently allow the system to find AdS or supersymmetric vacua. Ignoring these instabilities produces models that are superficially plausible, but ultimately inconsistent. Stabilizing the problematic directions in isolation is sometimes achievable, but doing so in a regime where all approximations are well-controlled is very challenging.

A similar story holds for supersymmetry breaking in
% last last call to separate susy breaking from flat/ds susy breaking
% June 19
string vacua with nonnegative cosmological constant.
Despite extensive efforts \cite{Silverstein:2007ac,Danielsson:2011au}, no
% Last call -- Liam  replaced -- -- with , so I can use -- -- a few lines down.
controllable, tachyon-free de Sitter solution is known for type IIA systems. In type IIB string theory, supersymmetry breaking by anti-D3-branes in a Klebanov-Strassler throat has been subjected to intense scrutiny \cite{Bena}.
% Last call -- Liam   substantial modification below.
The critical question is whether this breaking is encoded in purely normalizable perturbations -- and hence is dual to a metastable state of the cascading gauge theory, with energy controlled by the infrared scale -- or instead involves non-normalizable perturbations, and is dual to a perturbation of the gauge theory Lagrangian.  In the latter case, the scale of supersymmetry breaking would be high enough to preclude the use of anti-D3-branes for uplifting AdS vacua.
At the time of writing the evidence is rapidly evolving, but inconclusive.\footnote{See \cite{Dymarsky} for a candidate solution with exclusively normalizable perturbations.}
In other string theories, mechanisms for controllably small supersymmetry breaking
% last last call
% June 19 in flat and de Sitter geometries
in de Sitter space
are quite scarce, and this greatly hinders the construction of de Sitter vacua -- and cosmological models -- in these theories.

Overall, despite much tangible progress, no universally accepted and
%Liam June 10 {\it{calculationally tractable}}
calculationally tractable class of stabilized de Sitter vacua has been described to date, and building explicit compact models in which all moduli are stabilized within a domain of parametric control remains difficult. (The better situation for supersymmetric AdS and Minkowski vacua seems unhelpful, as de Sitter vacua, or Minkowski vacua with broken supersymmetry, are the preferred starting points for applications to the early or late universe.)
%Liam June 10 In particular, most
As a result, most
cosmological models developed in string theory to date do one or more of four things: they pay lip service to the problem of stabilizing moduli; they carefully invoke a method for stabilizing moduli without actually constructing a compact model with stabilized moduli;
they stabilize moduli in an explicit compactification, but rely on
%Liam June 10 imperfectly known
imperfectly-known
properties like higher-loop effects \cite{LVScosmo1,LVScosmo2}; and/or they uplift the ground state vacuum energy to flat or de Sitter values by adding antibranes \cite{KKLT,KKLMMT}.  The first approach is of little value, and the remainder carry some risk: the second presupposes that sectors associated with the unspecified regions of the compactification are cosmologically unimportant, despite ample evidence to the contrary; the third relies on plausible assumptions about higher loops that might turn out to be wrong; and the fourth is difficult to control convincingly \cite{Bena} (but see \cite{Dymarsky}).

One approach to the difficulty of
constructing a cosmological model using a particular string vacuum is to determine the most general possible form of the effective action describing an entire sector of interest. This can be achieved in sectors
corresponding to computable local geometries,
like Calabi-Yau cones probed by D3-branes: arbitrary effects of sources in the remainder of the compactification can be expressed in terms of Wilson coefficients of operators in the effective theory of the local geometry \cite{Baumann}. Although these coefficients could be computed, in principle, in a given compact model, contemporary methods are
%Liam June 10 as yet (redundant with contemporary)
inadequate for this task.  Despite appearances, this is not a dead end: in a Faustian bargain, the inflationary phenomenology of the resulting model turns out to have negligible dependence on the detailed properties of these coefficients \cite{Nishant}. The result is a class of well-motivated inflationary models that plausibly arise in stabilized string compactifications. Unfortunately, they are of limited use in constraining string theory: the models can be constructed precisely because they are statistically insensitive to ultraviolet physics.

\subsection{Initial conditions}

Perfect knowledge of the dynamics that arises in a general compactification of string theory would not suffice to make cosmological predictions: we also require a theory of the initial conditions. How urgent this is depends on the extent to which the desired cosmology relies on
%Liam June 10 starting out
%Liam June 10 in an unusual configuration.  with
very specific initial conditions. The requirement of special initial conditions is a particularly embarrassing situation when it arises in inflationary models, part of whose original motivation came from their promise to provide a robust dynamical explanation for the
exceptionally smooth
%Liam June 10 initial
conditions required in later epochs.

In special circumstances, most notably chaotic inflation \cite{chaotic}, one can argue that the effects of initial conditions are washed away along the approach to a dynamical attractor. Because the energies involved approach (and the field values exceed) the Planck scale, string theory is clearly the best available framework for studying the extreme conditions at the onset of chaotic inflation.  But our present understanding is inadequate: nearly everything that we
know about stabilizing moduli applies only for geometric compactifications that are large in string units, with correspondingly limited energies. Consequently no plausible proposal yet exists connecting the initial phase\footnote{It is important to distinguish the initial epoch with Planckian energy densities, where \cite{chaotic} argues that an inflating universe emerges from Planckian chaos,
from the more staid period of large-field inflation at GUT-scale energies that could give rise to the observed density perturbations.   The latter picture is plausibly realizable in flux compactifications \cite{largefield,LVScosmo2}.}
of chaotic inflation
to, {\em e.g.}, flux compactifications on controllably large Calabi-Yau manifolds. Indeed, any scenario for inflation in a string-derived effective theory with cutoff $M$ can at best push the question of initial conditions to times of order $1/M$. For this reason, inflationary models in
effective supergravity theories derived from string theory are limited to the question of `earlier conditions' rather than of initial conditions.

Several approaches to this problem have been considered. False-vacuum eternal inflation in a large landscape of vacua might eventually lead, via tunneling, to a later stage of evolution responsible for the observed universe.
%Liam June 10 repetition At present this
This picture actually appears to compound the initial conditions
problem: strenuous efforts to understand measures in eternal inflation (see the contribution by Freivogel to this volume) have not led to a broadly-accepted prescription.
%Liam June 10 At
Thus, at present the connection to an earlier phase of eternal inflation only adds new unknowns, plagued by poorly-controlled infinities.

\subsection{Bounces}

A second approach to initial conditions is to postulate a `bounce' that connects an earlier contracting phase to the present-day expansion. Such cosmologies are also much explored because an epoch of decelerated contraction during the
%Liam June 10 very-early
very early universe shares many of the attractive cosmological features of accelerated inflationary expansion (see the contribution by Lehners to this volume). The fundamental problem for any such approach is to engineer the bounce, {\it i.e.} to identify dynamics that is capable of reversing the universal contraction.  In the rare cases where this can be achieved, a more difficult challenge looms: generating primordial perturbations that -- after evolving through the bounce -- are consistent with observations.

Bounces can be classified as nonsingular or singular, depending on whether they occur while the universe is still large enough to be described by a low-energy effective field theory, or in a regime where this effective description fails.\footnote{We are only concerned with bounces of the Einstein-frame metric.  Bounces in which the scale factor in the Einstein-frame metric evolves monotonically while the metric of some other frame undergoes a bounce are less constrained. Although it is tantalizing that the metric along whose geodesics photons move -- and so to which most cosmological observations refer -- need not be the Einstein-frame metric (for example, if photons are localized on a brane), this observation has not led to a fully convincing bouncing cosmology.}  Both labels require clarification:
the
bounces termed nonsingular
in the literature are classically nonsingular given sufficiently homogeneous, isotropic initial conditions, while so-called singular bounces that leave the purview of effective field theory are usually imagined
(but not shown)
to be nonsingular in the full quantum gravity description.  We will examine these cases in turn.

When the universe is large at the bounce, the low-energy
%Liam June 10 Einstein equation requires
Einstein equations require a violation of the null energy condition \cite{nullenergy}. For  Friedmann-Robertson-Walker (FRW) spacetimes this is
easily seen from the
Einstein equations
%Liam June 10,
in the form
\begin{equation}
 \left( \frac{\dot a}{a} \right)^2 + \frac{k}{a^2} = \frac{8 \pi G}{3} \, \rho
 \quad \hbox{and} \quad
 \frac{\ddot a}{a} = - \frac{4\pi G}{3} \, (\rho + 3p) \,,
\end{equation}
where $k = 0, \pm1$ is the usual curvature parameter, $\rho$ is the energy density,
%Liam June 10 added a ,
and $p$ is the pressure. Since $\dot a = 0$ and $\ddot a > 0$ at the instant of the bounce, if $k \ne 1$ then $\rho \le 0$ and $\rho + 3p < 0$. (FRW spacetimes with $k=+1$ require only a violation of the strong energy condition.)  This is a grave problem: isotropic backgrounds violating the null energy condition suffer from instabilities,
either tachyons or ghosts \cite{Dubovsky}. Effective field theories admitting bounces have been proposed \cite{Creminelli:2006xe, Creminelli:2007aq}, using ghost condensates \cite{ArkaniHamed:2003uy} to supply the violation of the null energy condition. However, these models have also been criticized \cite{Kallosh:2007ad} on the grounds that they suffer from fatal vacuum instabilities. Furthermore, the prospects for having an ultraviolet completion for such an effective theory, violating the null energy condition, appear rather dim. Ghost condensate backgrounds spontaneously break
Lorentz invariance, 
%%LM Aug08 clarified why these constraints do not apply
and hence are not directly constrained by the analysis of \cite{Adams}, 
but, as explained in \cite{ArkaniHamed:2007ky}, these theories are incompatible with cherished principles of black hole and de Sitter thermodynamics, and hence are unlikely to be consistent at high energies.  In particular, there are no hints as to how a ghost condensate might arise in string theory.

Even granting a suitable ghost condensate, instabilities present a very severe problem for classically nonsingular bounces.
Small initial anisotropies grow during the contracting phase, leading -- given sufficient time -- to chaotic mixmaster behavior and an anisotropic crunch rather than a nonsingular bounce. It was argued in \cite{Creminelli:2006xe, Buchbinder:2007ad} that this problem is absent if the bounce is very brief, with the contraction lasting for approximately one $e$-fold.  However, this conclusion was recently criticized in \cite{Xue:2011nw},
which argues that extreme fine-tuning either of any initial anisotropy ({\em cf.\ e.g.}\ \cite{Lin:2010pf}), or of the potential driving a period of ekpyrotic expansion,
%Liam June 10 are
is necessary to prevent a singular crunch.
One conclusion is clear: it remains an open problem to describe a bounce that is classically nonsingular,
%Liam June 10
that
is not exponentially sensitive to initial anisotropy, and
%Liam June 10
that
generates a spectrum of primordial perturbations in agreement with observations.

The second category of bounces are those that occur at such small sizes that the low-energy, low-curvature $\alpha'$ expansion breaks down, so low-energy field-theory reasoning is invalid. In this case the Einstein equation need not be a good approximation, so the null energy condition need not be violated. In principle, this might turn the breakdown of the $\alpha'$ expansion from a bug into a feature. However, loss of the $\alpha'$ expansion also deprives us of our main calculational tools, and our understanding of
%Liam June 10 time-dependent, highly curved,
highly curved time-dependent solutions in string theory is extremely limited.  A common response to this impasse is to assert that string theory could yield a smooth bounce, and then connect the expanding and contracting phases of the effective description with boundary conditions that yield satisfying phenomenology.
No evidence yet exists that such an approach provides an accurate description of a bounce in string theory. Moreover, once one imposes a set of
{\em{ad hoc}} boundary conditions that are not rooted in a controlled calculation, it is fair to ask whether some other set would serve equally well, making the predictivity of these scenarios questionable.

What is required is an approach that completely avoids the weak-coupling expansions that are failing. Significant efforts have been made along these lines to understand singular bounces in M-theory, specifically the collision of the end-of-the-world `boundary' branes in heterotic M-theory \cite{Turok:2004gb}. These bouncing geometries are classically singular, with divergent curvature at the instant of the bounce. Moreover, M2-branes stretching along the collapsing interval between the boundary branes become light (and so cannot simply be integrated out) when the bounce occurs: in the ten-dimensional description, the heterotic string tension measured in Planck units tends to zero. Two important questions are whether semiclassical production of stretched M2-branes -- or of other species whose mass varies during the bounce --
%Liam June 10 backreact
backreacts enough to disrupt the classical evolution, and whether the curvature singularity can be resolved in a fully quantum description.

Some progress has been made on the former question: in a much-simplified model involving a product of Euclidean space and a compactified Milne spacetime with line element
\begin{equation}
 \exd s^2 = -\exd t^2 + t^2 \exd \theta^2 \,,
\end{equation}
the semiclassical evolution of wrapped M2-branes has been argued to be nonsingular.  Even though the M2-branes become light at the bounce, within the approximations of \cite{Turok:2004gb} only a finite density of M2-branes is produced as a result.  Although necessary, this is not yet sufficient for a controllable bounce. The burning question that remains, as recognized in \cite{Turok:2004gb}, is whether the bounce is nonsingular in the fully quantum-mechanical treatment. M-theory in itself provides no guidance, because the low-energy supergravity normally used to describe it is inapplicable, and the full quantum theory is unknown.

In this sense, it would be misleading to say that M-theory in its present form provides a framework for studying bounces: as soon as the problem becomes interesting, the low-curvature expansion becomes invalid.  Heterotic string theory could in principle provide a consistent description of the moment of collision, but although the string coupling vanishes at the bounce, the $\alpha^{\prime}$ expansion fails completely.  Describing the collision of the end-of-the-world branes in a computable regime remains a distant --- yet worthwhile --- goal, and it is not
clear that the outcome is a nonsingular bounce.

\subsection{Singularities in string theory}

Although M-theory may not yet allow a precise treatment of the singularities arising in bounces, string theory is not completely silent about the problem of singularities. Indeed, an important motivation for applying string theory to cosmology is the hope of understanding the Big Bang singularity, either by exposing its microphysical structure or replacing it with a bounce connecting to a prior contracting phase \cite{SGcosmo, Veneziano}.

As emphasized above, this is a difficult row to hoe, since it is insufficient to work within a low-energy effective theory derived from string theory.
For a generic spacelike singularity -- including the most interesting case of the Big Bang singularity in an FRW cosmology -- this obstacle has proven to be an impasse: string theory provides hardly any illumination at present. However, in simpler toy models there has been considerable progress.

An important class of examples in which string calculations are possible are
null\footnote{Timelike singularities are so different from spacelike ones that an understanding of the former is likely to provide only the vaguest guidance to the nature of the Big Bang.} singularities in Lorentzian orbifolds \cite{Horowitz, Balasubramanian:2002ry, Cornalba:2002fi, Liu:2002ft}. These models are described by exactly solvable CFTs, and so can evade the threat of a breakdown of the $\alpha^{\prime}$ expansion.
%Liam June 10 about a classical background.
Even so, string scattering amplitudes in these backgrounds exhibit divergences \cite{Liu:2002ft} which can be attributed to the formation of large black holes \cite{Lawrence:2002aj, Horowitz:2002mw} after the introduction of even a single particle.  This would seem to preclude the use of these geometries as toy models for the Big Bang.

In exact plane waves in $d+2$ dimensions of the form
\begin{equation}
\exd s^2 = -2 \exd X^{+} \exd X^{-} + \sum_{i=1}^d \Bigl[ - F(X^{+})
(X^{i})^2 (\exd X^{+})^2 + (\exd X^{i})^2 \Bigr]   \,,
\end{equation}
the background receives no $\alpha^{\prime}$ corrections \cite{Horowitz2}, even when $F(X^{+})$ is such that the background contains a null singularity, and one can quantize strings explicitly \cite{Papadopoulos:2002bg}. It is then interesting to inquire about the behavior of strings that encounter this singularity. As stressed in \cite{Craps:2008bv}, a consistent approach to the description of strings in the background of the null singularity involves quantizing the strings and performing the calculation of interest in a resolved geometry, {\em i.e.}\ with $F(X^{+})$ chosen such that the singularity is absent, and then examining the singular limit.

Of course, the possible resolutions are legion, and without a well-motivated prescription for selecting one, we are back to square one in terms of characterizing the physics of the singularity.  There is a unique resolution of the singular exact plane wave that does not introduce new dimensionful parameters hidden in the singularity, but with this scale-invariant resolution the energy of a free string diverges upon crossing the singularity \cite{Craps:2008bv}. Three unpalatable alternatives remain: either the singularity is impassable; a resolution beyond our ken introduces new scales into the problem; or a miracle arises in the interacting theory.

Null Kasner-like geometries can also be used as models for the near-singularity region of a more general spacetime  with a  null singularity \cite{Madhu:2009jh}, and the free bosonic string can be quantized in these backgrounds.  The region near the singularity turns out to be awash in highly-excited strings, raising the question of backreaction in the interacting theory.

Quantizing the free string in an exact background is not the only way to use string theory as a window onto singularities.  Spacelike and null singularities in asymptotically AdS spaces can be described via AdS/CFT, with the weakly-coupled gauge theory providing the (in principle) computable quantum-mechanical description of the would-be singularity. There is suggestive evidence that for spacelike singularities, the evolution of the CFT dual does not remain well-defined for all time,
implying that even in the full string theory, the evolution ends with a big crunch \cite{Hertog:2004rz}.  Null singularities in $AdS_5$ are somewhat better behaved, and the gauge theory dynamics described in \cite{Das:2006dz, CHT} is nonsingular, even though the string oscillator modes become light near the singularity.

The upshot is a familiar one: string theory provides remarkably powerful theoretical tools that equip us to study, in intricate detail, a range of scenarios that do not correspond to the observed universe.
Thus, although some promising approaches to cosmological singularities in string theory exist, it remains much too early to declare victory.

\subsection{String-inspired cosmology}

% Liam Why is this "despite"?
%Liam June 10  still stuck on the right introductory phrase here
Despite the above difficulties, it is a common lament that effective field theories
derived from string compactifications
are inadequately `stringy', and one should preferentially pursue approaches to cosmology that rely more heavily on the structure of string theory. After all, if we always seek the implications of string theory where things look fairly mundane --- like a higher-dimensional supergravity --- then we should not be surprised to find mundane results. We risk missing string theory's most insightful predictions.

It is quite possible that intrinsically stringy cosmology might be interestingly different
%Liam June 10 than
from that coming from field theory.
%Liam June 10 After all, even
Even something as simple as a thermal bath of strings in a
%Liam June 10 flat,
toroidal spacetime can differ significantly from a thermal gas of particles, since the ability of strings and branes to wrap themselves around the toroidal directions gives rise to the T-duality that ensures equivalence between toroidal radius $r$ and radius $\alpha'/r$.
Naively applied in a cosmological context, this suggests a bouncing cosmology wherein at present the temperature $T$ falls as $r$ grows, but in the
%Liam June 10 hyphenation
`trans-Planckian past' $T$ grew as $r$ increased (and so the
%Liam June 10 `dual'
dual size $\alpha'/r$ decreased) \cite{SGcosmo,Veneziano}.

Although
%Liam June 10
this is
a very suggestive picture, the devil is in the details, and it remains a distant goal to solidly base such a cosmology on controlled string calculations. To understand the system at all one truncates the
particle content to a lower-dimensional
subset of states invariant under T-duality --- typically the metric, the dilaton and possibly a few other lower-dimensional fields --- and seeks cosmological solutions to their equations of motion in the presence of stress-energy
whose properties are
motivated by a toroidal T-dual string gas. The basic difficulty is
%Liam June 10 a familiar one:   phrase used on preceding page
familiar: since
%Liam June 10 to the extent that
semiclassical calculations with this field theory
%Liam June 10 are regarded
provide only a low-energy approximation to the evolution of the full string system,
%Liam June 10 it is
they are inadequate for inferring how spacetime responds to the enormous energy densities encountered in the self-dual regime where contraction converts to growth.

To the extent that this field theory is simply regarded as a truncation of the full theory, rather than
%Liam June 10
as
a low-energy approximation, it is no longer clear why the evolution of the truncated system should agree with the evolution of the real string system. In particular, once time derivatives are as large as the string scale, there is no reason why the evolution should be adiabatic, and (among other things) one should ask why states associated with previously truncated fields are not produced anew by the rapidly evolving conditions.

% last call 2: split this off and inserted a line about mirage and brane cosmologies
At present, cosmological models built along these lines
%Liam June 10 must be
are best
%Liam June 10 regarded
described as being string-inspired, rather than as {\em bona fide} consequences of string theory. The same can also be said for other approaches that use string-motivated ideas but are only loosely based on constructions using specific string vacua. These include `mirage' cosmologies \cite{mirage}, for which a time-dependent metric is obtained for observers trapped on a moving surface that moves through a static extra-dimensional spacetime, as well as a broader class of `brane-world' cosmologies. What is taken from string theory in these models is the idea that observers can be localized on surfaces, just as might happen if these observers were constructed from open strings attached to D-branes. However, the properties of these surfaces are typically chosen for the convenience of the cosmology rather than from explicit string calculations using D-branes.

Of course, string-inspiration in itself need not be a bad thing, depending on one's goals. Inspiration is good if it can identify new mechanisms for solving low-energy problems, and new ways to think about cosmology. Using string theory as a muse in this way can be useful. In particle physics this kind of reasoning has identified several new mechanisms for solving the gauge hierarchy problem by lowering the
scale of gravity \cite{braneworld,ADD,intscale,RS}.

The downside of inspiration is that string-inspired models may well be unreliable indicators of how string theory actually behaves in a cosmological setting. In this sense string-inspired models could be as different from string cosmology as medicine-inspired treatment is from medicine.

\subsection{The string-compactification inverse problem}

Even imagining that observations give us important new windows into the properties of the very early universe, {\em e.g.}\ through a detection of non-Gaussianity or primordial gravity waves, what does this tell us about string theory?
% last call 2: changed this next line
Although any such a detection would transform our picture of how inflation works -- telling us much about the energy scales responsible for inflation and the mechanism responsible for generating primordial perturbations -- unfortunately, with the present state of the art, we learn very little about string theory itself.

Barring unusual good fortune -- like the discovery of a cosmic network of $(p,q)$ strings \cite{pqstrings} -- observations are likely only to point us towards a family of four-dimensional effective field theories, involving the four-dimensional metric coupled to some number of (possibly scalar) matter fields. This situation would prompt the following three questions: can these effective theories arise in string theory? If they do, how does this constrain the allowable compactifications?  Most importantly, can these inferences be used as evidence for or against string theory?  Regrettably, the present answers are likely to be `probably', `not significantly', and `no', respectively, as we now explain.

On the first two points, although one can in some cases derive the leading terms in the effective
%Liam June 10 Lagrangians
Lagrangian for a specific string compactification, we are just now beginning to explore the space of stabilized vacua.  Failure to find a desired effective theory in a search of currently-known string compactifications sheds little light on the prevalence of this effective theory in the full space of solutions of string theory. Moreover, we have no measure even on the known subspace of the space of compactifications, so constraints have little bite.

Turning to the question of constraining string theory itself, a fundamental difficulty is the ubiquitous crutch of {\em four-dimensional} effective field theory (as opposed just to effective field theory), which underpins nearly all calculable dynamics in string cosmology. Four-dimensional field theory is used because it is much easier to solve for the homogeneous time-dependence of the effective four-dimensional geometry in the presence of time-dependent four-dimensional scalar moduli than it is to compute the time-dependence of the full, higher-dimensional geometry. Indeed, the existence of classical solutions to higher-dimensional theories having de Sitter curvature in four dimensions is constrained by increasingly strong
%Liam June 10 `no-go'
no-go theorems \cite{dSnogo}
%%CB:Aug01: added this because it is an example of a dS solution in a higher-dimensional supergravity that evades these no-go theorems (although one without a known stringy provenance)
(see, however, \cite{6DdS}).

As a consequence of this reliance on the effective description, the predictions of string cosmology are couched in terms of the four-dimensional effective action that is believed to apply. The predictions are then by construction indistinguishable in principle from four-dimensional field theory results. The most one might hope is that the Lagrangians that are natural in string theory are not natural in traditional phenomenological model building in field
theory.\footnote{We will not address here the related question of whether the field theories of interest for cosmology might lie in the `swampland' --- {\em i.e.} be effective theories for which there is no ultraviolet completion in string theory \cite{swamp} --- because our present tools do not allow a clear delineation of the swampland's boundaries.}
%%CB Aug01: added this footnote about the swampland.
%%LM Aug08: minor rephrasing
 
This hope is not without merit. One thing that appears generic in contemporary string compactifications is the presence of multiple light fields, more than one of which can be cosmologically active. This happens because there are many degrees of freedom, and the mechanisms that make one of them light enough for cosmology tend also to make others light. By this line of reasoning, the generic effective cosmology arising in string theory would involve multiple dynamical fields.  In Calabi-Yau compactifications, the number of active moduli ranges roughly from a few to a few hundred, and the upper end of this range is decidedly unnatural -- though certainly consistent -- in purely four-dimensional model building.

Cosmological observations could very readily provide firm evidence for more than one light field in the early universe, for example by revealing non-Gaussianity that is large in the squeezed limit: any single-field model of inflation, regardless of slow roll, predicts negligible non-Gaussianity in this regime \cite{Maldacena, Zaldarriaga}.  Further theoretical developments would be required to tease out evidence for {\it many} light fields: the dynamics becomes intricate, the observables proliferate, and analytic methods typically fail.  Even in the two-field case, the predictions depend sensitively on the details of the evolution of the universe between earlier epochs and now.  As a result, one cannot now say much that is incisive, and understanding cosmological evolution in the presence of many light fields remains a very interesting challenge for the future.

\section{A way forward?}

Having wrung our hands over the imperfect state of string cosmology at present, we now offer a few suggestions about how progress might nevertheless occur.

The most conservative approach is to work in computable regimes of explicit string compactifications and develop well-controlled inflationary models in the resulting four-dimensional effective theories.  This is akin to setting up shop directly under a well-lit lamppost.  All results are verifiable, and the striking success of inflation as a phenomenological paradigm makes this a safe bet.  One obvious criticism is that this approach leaves unexplored a great unknown; a more immediate concern is that hardly any examples with this degree of computability have been exhibited, despite years of work by many hands.

A slightly more risky strategy is to obtain effective theories from string theory by constructing models whose ingredients definitely exist in isolation, without exhibiting a compact space with all the desired properties.  One then argues that a compactification containing all the required ingredients plausibly exists.  The weaknesses of this approach are that plausibility is subjective, and one can easily overlook new sectors or interactions that are necessarily present in real stabilized compactifications.

Less useful is to write down effective theories with some set of attractive properties and assert, without calculation and against all contrary evidence, that these theories arise in string theory. Although model-building
that is consistent with the rules of effective field theory is above reproach,
a tenuous connection to string theory should not be used to paper over gross violations of these rules. As mentioned earlier, what value there is in this approach lies in any novel mechanisms that might emerge, and subsequently be more precisely justified.

Most useful are mechanisms to deal with the naturalness problems that are generic
%Liam June 10 to any
in systems with light scalars (which are so useful in cosmology), keeping in mind that the naturalness problems of cosmology are mostly about those interactions that are {\em not} written down when defining a model, rather than those that are. These problems are not solved simply by a loose appeal to a string theory connection.

In a nutshell, reliably anchoring cosmology in string theory is hard, and success is rare. Most current examples advertised as string cosmology invoke string theory, branes, or extra dimensions, but cannot be precisely identified as quantitative approximations to solutions in string theory, in D-dimensional gravity, or in field theory, and hence are better described as string-inspired.
Relatively few
scenarios are founded on controlled approximations around stabilized vacua, even if one permits optimistic assumptions about the domain of validity of the expansions involved.
Moreover, at the time of writing, all such examples are computable inasmuch as they remain within the domain of effective supergravity.

%Finally, the models that are constructed explicitly in string theory, are consistent with all observations,
%and can be distinguished from cosmological models in effective field theory
%can be counted on one hand (with no fingers).

The obstacles that we have described are forbidding but not insuperable, and a successful grounding of cosmology in string theory remains a worthy and realistic goal.
Our view is that progress is most likely to spring from searches that
\begin{itemize}
\item stabilize all moduli in explicit compactifications;
\item deal carefully with naturalness issues;
\item incorporate the effects of heavy fields on the cosmological evolution of light fields;
\item work within well-founded approximation schemes;
\item are forthright about calculational limitations and questionable assumptions; and
\item focus on the aspects of cosmology that are most sensitive to ultraviolet physics.
\end{itemize}

\section*{Acknowledgements}
% last call: added to the acknowledgements
We thank D.~Baumann, R.~Brandenberger, and F.~Quevedo for comments on a draft.
The research of L.M. was supported by the Alfred P. Sloan Foundation and by the NSF under grant PHY-0757868. C.B.'s research is supported in part by funds from the Natural Sciences and Engineering Research Council (NSERC) of Canada. Research at the Perimeter Institute is supported in part by the Government of Canada through Industry Canada, and by the Province of Ontario through the Ministry of Research and Information (MRI).

\end{document}